\documentstyle[onecolumn,psfig]{mn}

\def\lesssim{\,\lower2truept\hbox{${<\atop\hbox{\raise4truept\hbox{$\sim$}}}$}\,}
\def\gtrsim{\,\lower2truept\hbox{${> \atop\hbox{\raise4truept\hbox{$\sim$}}}$}\,}

\def\ref{\par\noindent\hangindent 20pt}

\begin{document}

\title{Mass function of the dormant black holes and 
the evolution of  the Active  Galactic Nuclei
}

\author[P. Salucci et al.]
{Paolo Salucci$^1$, Ewa Szuszkiewicz$^{1,2}$, 
Pierluigi Monaco$^3$ \& Luigi Danese$^1$ \\
$^1$ International School for Advanced Studies, SISSA, Via Beirut 2-4,
I-34013 Trieste, Italy\\
$^2$ Astronomy Group, Dept. of Physics and Astronomy, University
of Leicester, University Road, Leicester LE1 7RH, UK \\	  
$^3$ Institute of Astronomy, Madingley Road, CB3 0HA Cambridge,
UK \\	}

\date{}

\maketitle  

\markboth{Dormant BH and AGN}{}

\begin{abstract}

Under the assumption that accretion onto massive black holes powers 
active galactic nuclei (AGN), the mass function (MF)
of the BHs responsible for their past activity is estimated. For this, 
we take into
account not only the activity related to the optically selected AGN,
but also that   required to produce the Hard X--Ray Background (HRXB).
The MF of the Massive Dark Objects (MDOs) in nearby quiescent galaxies
is computed by means of the most recent results
on their demography.
The two mass functions match well under the assumption that the
activity  is concentrated in a single significant burst
with $\lambda=L/L_{Edd}$ being a weakly increasing function of luminosity.
This behaviour may be indicative of  some level of recurrence  and/or
of accretion rates  insufficient to maintain the  Eddington 
rates  in low luminosity/low redshift  objects.
Our results  support the scenario  in which the early phase of intense
nuclear activity occurred mainly in early type galaxies (E/S0) 
during the relatively short period in which they had  still an abundant 
interstellar medium. Only recently, with the decline of the
QSO luminosities, did the activity in late type 
galaxies (Sa/Sab) become statistically significant.
 
\end{abstract}

\section{Introduction}

The number counts of AGN's  and the intensity of the backgrounds
at high energies show that the  activity in nuclei of galaxies 
was much higher in the past than in the local universe. If we accept the 
paradigm that nuclear activity in galaxies is sustained
by accretion onto  massive BHs (see Rees 1996 for a review), then  
the problem of the location and discovery of the remnants of such
past activity immediately arises. There is evidence for the
presence of Massive Dark Objects (MDOs) in the centers of 
most, if not all, nearby galaxies with  a large spheroidal component.
The mass  of  MDOs  has been evaluated by using  very high resolution
spectroscopy and photometry of the centers of  nearby host galaxies.
Observations with HST have allowed a significant breakthrough in the 
angular resolution and have led to an enormous increase in sensitivity 
and precision of mass estimates
(Magorrian et al.\ 1998, hereafter M98;
van der Marel 1997, hereafter VdM97; Ford et al 1997, hereafter F97; 
for a review, see Kormendy and Richstone
1995, hereafter KR95). As a result, a large number of MDOs have 
been detected suggesting that we
are discovering the fossil of the past nuclear activity.
Here we will assume that MDOs are the BHs now dormant after
a shining past. 

The main  purpose of this paper is to  show that the mass function 
of the BHs we infer from the observations of the past activity
of AGN and QSOs corresponds to the  MDO mass function of the
local non active galaxies.

M98 have successfully exploited the very high 
resolution of HST photometry and ground based
spectroscopy of 36 E and S0 galaxies in order to estimate
their MDO masses. In spite of the large scatter in the data,
 they also  confirmed the existence of correlation between
the mass of the hot galactic component $M_{sph}$ 
and the MDO/BH mass, $M_{MDO}$, already suggested by Kormendy (1993) 
and KR95 and successively claimed  by M98 and 
van der Marel (1998) (hereafter VdM98,  by analysing
a sample of 46 early 
type galaxies).
These large samples, together with several  smaller ones
(e.g. F97; VdM97; H98) allow to estimate the distribution
function of the ratio $M_{MDO}/M_{sph}$ and then to evaluate
the mass function of the MDOs (see section 2.1).

A different approach to  evaluate the  MDO/BH mass function relies
on the hypothesis that radio emission from the nuclei of 
radio quiet galaxies is related to the mass of their MDOs.
As a matter of fact, a strong correlation
between MDO masses and nuclear radio luminosities has been
found by Franceschini et al (1998). This correlation, in connection
with the radio LF of the  nuclear 
emission of radio quiet galaxies, can be used to probe 
the MDO mass function (see section 2.2). 

In section 3 we will  determine the
mass function of the material accreted onto  BHs 
by exploiting the knowledge on the evolution of AGN/QSO LFs.
Reliable luminosity functions and cosmic evolutions
are presently available for optically  and soft X--ray selected objects.
On the other hand, we must also consider  a class of  heavily absorbed AGN,
under-represented in optical 
and soft X--ray surveys, that shows up in hard X--ray surveys.
These absorbed AGN are the most likely 
contributors of a major portion of the intensity
of the 2-50 keV X--ray background (HXRB) through BH accretion 
energy output. 
In fact,  optical identifications of 
serendipitous sources detected by BeppoSAX in the 5-14 keV band
 show that absorbed AGN significantly contribute to the HXRB 
(Fiore et al.\ 1998). Moreover, independent observational evidence 
supports unified schemes for the AGN, whereby
a large fraction of absorbed (type II) AGN is expected
(see Antonucci 1993 for a comprehensive review).
Thus, we will derive the mass function of relic BHs  by  including
not only the mass accreted on type I or optically selected AGN's, but 
also that accreted on type II AGN's.

In section 4
the comparison of the MF of the dormant BHs to the accreted
MF will be used to cast light on the characteristics
of the evolution of the nuclear activity in galaxies
of different morphological type.

We will adopt $H_\circ=70 km/s \ Mpc^{-1}$ unless  otherwise stated. 
Moreover,  $h=H_{\circ}/$ in units of $  100\  km s^{-1} Mpc^{-1}$. 

\section{Estimates of the MDO mass function}
 
\subsection{From the optical luminosity function to the MDO mass 
function}

Observations and subsequent analysis have
shown that MDOs are
quite common in galaxies with significant spheroidal components 
and that their masses are correlated with the spheroid masses, 
though with large scatter (Kormendy 1993, KR95;
VdM97; VdM98; M98; F98). 
The mean value
of $M_{MDO}/M_{sph}$ is still under debate and it ranges from 
$\langle M_{MDO}/M_{sph}\rangle \sim 10^{-2}$ (M98) to $
\langle M_{MDO}/M_{sph}
\rangle \sim 2\times 10^{-3}$ (H98). 
The uncertainty  is related to different assumptions
on dynamical models, particularly on the two-integral phase--space
distribution function (VdM97; H98; M98). 
Exploiting very high resolution HST photometry and ground based
spectroscopy, M98 estimated the mass of 36 MDOs in the centers
of nearby galaxies, mainly E and S0s. They found that the distribution  
of the ratio $x=M_{MDO}/M_{sph}$ can be described by a  Gaussian distribution in $log \ x$:

\begin{equation}
f( log \ x)=N \exp \ \Big(-{1 \over 2}(log \ x-log \ x_o)^2/\sigma^2\Big)
\end{equation}
where $N$ is a normalization constant.
The best fit values found by M98 are $log \ x_o\equiv log\langle 
 \ M_{MDO}/M_{sph}\rangle  =-2.28$
and $\sigma=0.5$.
Analyzing the very high resolution photometry of 46 early type galaxies
and  reproducing it by means of the  models proposed by Young (1980),
VdM98 found $log\ M_{MDO}\simeq -1.83 + log \  L_V$ 
with rms scatter $\sigma \sim 0.3$ dex. 
Using standard values of
$M/L_V\simeq 5.5\ h(L_V/L_{\star})^{0.25} $ for spheroids
(see below), this corresponds to 
$\langle log \ M_{MDO}/M_{sph}\rangle \simeq -2.64$. 

%%%%%%%%%%%%%%%%%%%%%%%%%%%%%%%%%%% FIG 1new %%%%%%%%%%%%%%%%% 
\begin{figure}
 
\vspace{7.8cm}
\includegraphics{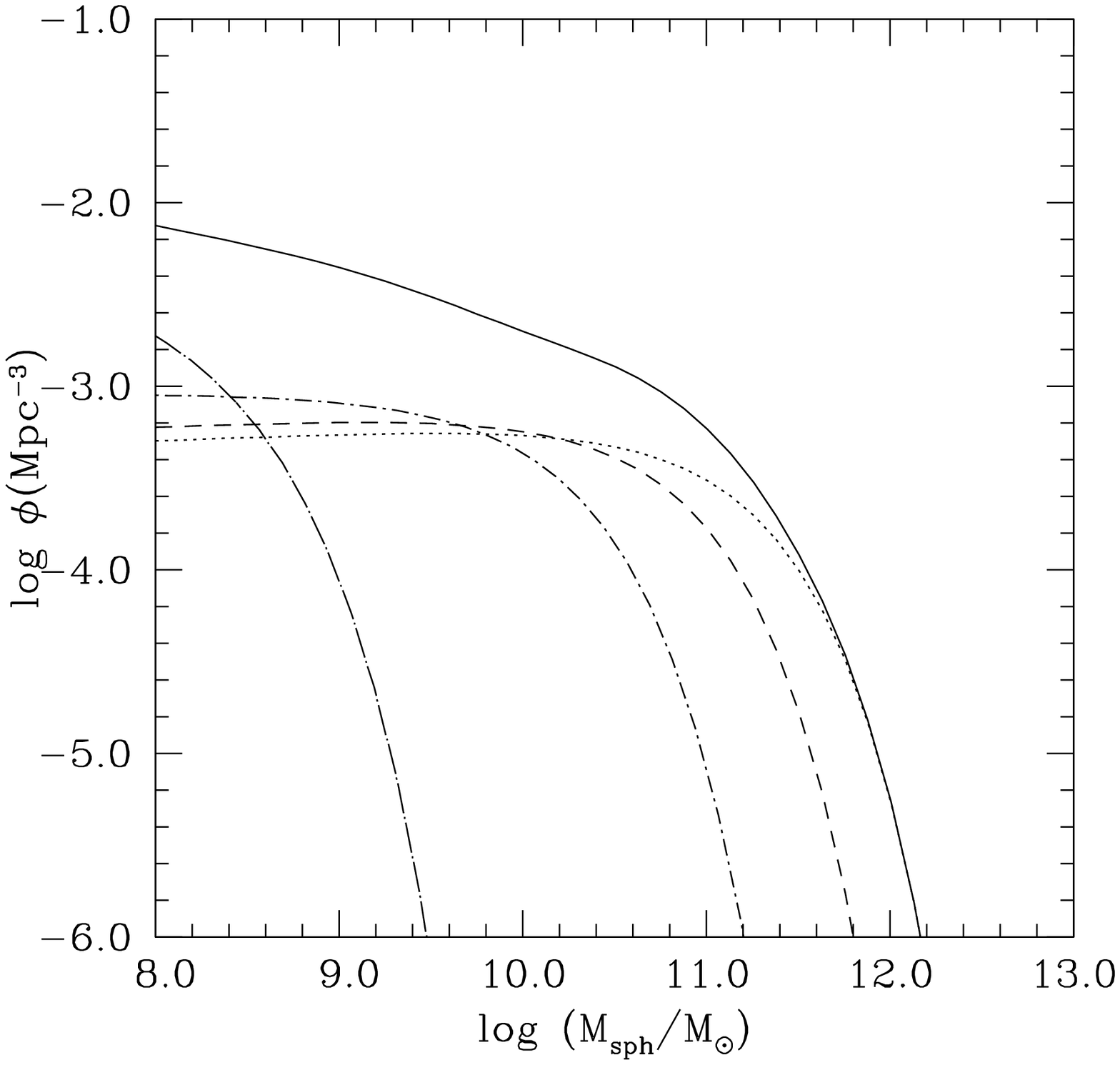}         
\includegraphics{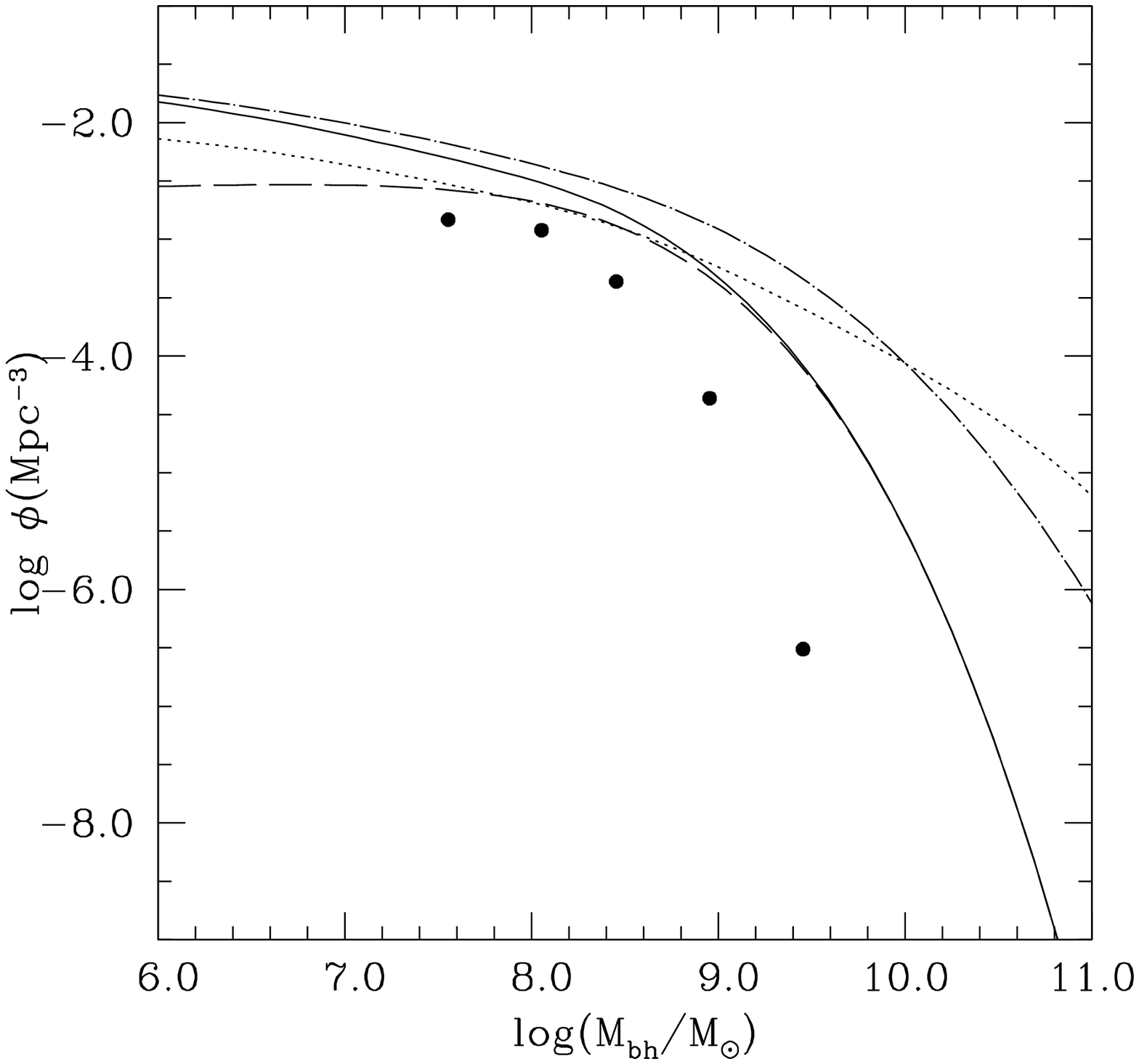}         
\caption{{\it Left panel: a)} The mass function of the spheroids.
The solid line is the total MF, the short-dotted, dot-dashed, dashed and 
dot-long dashed lines refer to 
spheroids in E/S0/Sab/Sbd.
{\it Right panel: b)}
The MDO mass function, $\phi_{OMF}dlog \ M_{BH}$:
{\it i)} derived from 
eq (1) with the values of  $\langle x\rangle$ and $\sigma$
given in section 2.1 (solid line). 
{\it ii)} relative to  the  contribution of MDO's  
in E/SO's  (long dashed line) {\it iii)} derived by
 Franceschini et al (1998) (solid circles)
{\it iv)} derived by   M98 (dotted line).} 
\end{figure}
%%%%%%%%%%%%%%%%%%%%%%%%%%%%%%%%%%%%%%%%%%%%%%%%%%%%%%%%%%%%%%%%%%

We have constructed a sample of 30 galaxies, for which at
least two independent mass estimates of their MDOs are
available in the literature (KR95, VdM97, VdM98, H98, F98, M98).
Taking the geometric mean as the best estimate of the ratio
$x$, we have found that a Gaussian  distribution in $log \ x$,
with $\langle log \ \ x\rangle =-2.60$ and  
$\sigma=0.3$ is a good description of the
data. This result, quite close to that obtained by
VdM98 and  in agreement with the estimates of  H98 and F98, 
will be adopted  as a reference distribution. 

The estimate of the spheroid mass function in the
local universe is a  preliminary step in  the evaluation 
of the  MDO MF. We determine the luminosity function
of the spheroids by using the total-luminosity
function ($L_{tot}$) and the estimates of the ratio of the spheroidal
to total luminosity $L_{sph}/L_{tot}$
for  different types of galaxies (E, S0, Sa/Sab, Sbc/Scd).
The LFs of early type and
spiral galaxies are well known in the range
$-23\leq M_B\leq -17$. In fact, results 
of different groups (e.g. Efstathiou et al., 1988;
Loveday et al.,  1992;  Zucca et al.,  1997; 
Geller et al.,  1997,  Heyl et al.  1997), based on different 
galaxy samples, well agree in the luminosity range relevant to our
purpose.  In order to represent the LFs of the various Hubble  types,
we use standard Schecther functions whose  parameters $L_*$
and $\alpha$  are given in Table 1
for  $h=1 $  

\begin{table*}
\begin{tabular}{lcccc}
Galaxy type & $\Phi_{\star} h^3$ & $M_B^{\star}+5 log \ h$  & $\alpha$ & $f_{sph}$ \\
       & $Mpc^{-3}$            & $ $  &    & \\
 (1)& (2)&(3)&(4)&(5)\\
\hline
E     & $2.1\times 10^{-3}$ & $-20.$ &  $-0.95$& $0.9$   \\
S0    & $2.4\times 10^{-3}$ & $-19.6$ &  $-0.95$& $0.65$  \\
Sab & $2.9\times 10^{-3}$ & $-19.8$ &  $-1.0 $& $0.40$  \\
Sbc/Scd & $7.6\times 10^{-3}$ & $-19.3$ &  $-1.3 $& $0.10$  \\
\end{tabular}
\end{table*}

The ratio $L_{sph}/L_{tot}$ has been investigated by several authors
(Simien and de Vaucouleurs 1986, Kent 1985, Kodaira
 et al. 1986) and it has been found to vary along the Hubble sequence.
There is a reasonable agreement 
on the average values, though intrinsic scatter is significant,
particularly in the late types. 
With the  values  reported in Table 1,
the total luminosity
density in spheroids turns out to be $\sim 30\%$ of the total
luminosity density, in agreement with Schechter and Dressler (1987)
and Fukugita et al.\ (1998).

From the spheroid LF we  infer  the MF 
by adopting $M_{sph}/L_V\simeq 5.5\ h(L_V/L_{\star})^{0.25} $, corresponding
to $M_{sph}/L_B\simeq 6.9\ h (L_B/L_{\star})^{0.25} $ with (B-V)=0.90.
This law is compatible at 2 $\sigma$ level with
$(M_{sph}/L_V)\propto L_V^{0.18\pm 0.03}$ found by M98 and 
$(M_{sph}/L_B)\propto L_B^{0.35\pm 0.05}$ claimed by van der Marel (1991);
consistent with stellar evolution models. The spheroid
MF is presented in figure 1a. The overall shape of the MF
is relatively flat at $M_{sph} \lesssim 10^{11} M_{\odot}$
and exhibits an exponential decline for $M_{sph} >10^{11} M_{\odot}$.
At $M_{sph} > 5\times 10^{10} M_{\odot}$
the mass function is dominated by spheroids in E and S0 galaxies.
The Sa/Sab galaxies exhibit an exponential decline of their
MF for $M_{sph} > 10^{10} M_{\odot}$ and Sbc/Scd galaxies for $M_{sph}> 10^8
M_{\odot}$.

The  MDOs mass function (hereafter OMF)  has been computed by
convolving the MF of the spheroids with the 
adopted log Gaussian distribution with $\langle M_{MDO}/M_{sph}\rangle =-2.6$
and $\sigma=0.3$.
As it is apparent in figure (1b), the scatter of the ratios
$x=M_{MDO}/M_{sph}$ softens the exponential fall off of the luminosity
function;  the broader the distribution
the gentler is the decline.
Of course, the convolution relies on the assumption that scatter 
reflects a  real complexity 
of the physical processes
leading to the formation of the BHs in galaxy centres. 

The total mass density amounts to
$\rho_{MDO}\simeq 8.2 \times 10^5\ M_{\odot}/Mpc^3\ h^2$, with
large fraction $\sim 75\%$ due to MDOs in E and S0 galaxies.
Of the remaining $20\%$ is  due to MDOs in Sa/Sab galaxies and $5\%$  
to MDOs in late type spirals.
The results for MDOs in spirals are consistent with the upper limits
that Salucci et al.\ (1998) have derived from the analysis
of a large number of high quality rotation curves. 

The differences with the MFs derived
using the distributions proposed by M98
are apparent from figure 1b. The M98 log Gaussian law
predicts  a large number density of BHs with mass larger than 
$10^{10} \ M_{\odot}$ and 
yields a  total mass density $\rho_{MDO}\sim 3\times 10^6\ M_{\odot}/Mpc^3
\ h^2$, which is higher than the
estimate of the local mass density of the accreted matter,
based on counts of  optically selected AGN,
by a factor of about 10--15 (see section 3).

A contrasting result has been obtained by Franceschini et al.\ (1998),
whose mass function, estimated from the optical LF
of E/S0 galaxies and the correlation between $M_{MDO}$ and the
luminosity of the bulge of the host galaxies,  is substantially
below our determination and yields a lower total mass
density $\rho_{MDO}\sim  10^5\ M_{\odot}/Mpc^3
\ h^2$. This is partly due to the adopted LF of E and S0
galaxies and partly to the fact that they neglected the
broadness of the distribution of the ratio $M_{MDO}/L_{sph}$.

\subsection{From the radio luminosity function to the MDO/BH mass function}

A different but complementary approach to the evaluation of the MF of BHs
is based on the idea that the radio emission from cores of normal galaxies
is related to the mass of the hosted inactive BHs. Actually,
accretion onto dormant BHs occuring
at very low rates  is possibly 
advection--dominated with low radiative efficiency
(Rees et al 1982; Abramowicz et al 1988; Narayan
and Yi 1995 a,b).  In this
regime, a hot ion torus ($T\sim 10^{12}\ ^{\circ} K$) forms, while the
electrons, coupled to the ions only through two--body Coulomb effects,
remain at much lower temperature ($T\sim 10^{9}\ ^{\circ} K$) and emit at radio
frequencies via thermal synchrotron. 
 Fabian and Rees (1995) pointed out that the thermal synchrotron emission 
 of the electrons might be detectable in cores of quiescent galaxies.
 They also noted that most of the elliptical and
 S0 galaxies exhibit a low--power emitting radio core. 
 The ADAF model predicts for the
 core radio power $P_{\nu}\propto \nu ^{2/5}\dot  M^{4/5}M_{BH}^{2/5}$
 (e.g. Mahadevan 1997).
 The dependence on frequency is quite close to the relationship
 $P_{\nu}\propto \nu ^{1/3}$ found by Slee et al (1994)
for a sample
 of radio cores of E/S0 galaxies. 
 However the ADAF model should be taken with some caution, since
new high resolution radio and sub--millimeter
observations of three giant elliptical galaxies significantly disagree
with its canonical predictions, although 
the possibility of explaining the observed spectra with
modifications of the canonical ADAF is not ruled out (Di Matteo et al.\ 1998).
With Bondi accretion rate
$\dot M\propto M_{BH}^2 \rho(\infty)/c_s^3(\infty)$ 
the ADAFs yield radio powers
$P\propto M_{BH}^{2.0}$, if the  density and sound velocity
at the boundaries of the accretion flow are independent
of the mass of the spheroid. 
When the latter depend on
spheroid mass, $\rho \propto M_{sph}$ and
$c_s \propto M_{sph}^{1/4} \propto M_{BH}^{1/4}$, we have
$P\propto M_{BH}^{2.2}$.
Given that self absorbed processes (e.g. jets)
yield $P\propto M_{BH}^2$,  
a relationship $P\propto M_{BH}^{\alpha}$ with
$2.0 \lesssim \alpha \lesssim 2.2$ is expected under rather
general conditions.
 
Franceschini et al.\ (1998) find a 
significant correlation between the radio power of the cores  $P_{core}$ 
and the estimated BH mass in 8 objects
with $P_{core}\propto M_{BH}^{2.7\pm 0.6}$. In a new
analysis, we searched  for high angular resolution radio observations of  
the galaxies with at least two 
MDO mass estimates, as specified in the section 2.1.
The radio observations reported by
Wrobel (1991), Wrobel \& Heeschen (1991),
Slee et al.\ (1994), Sadler et al.\ (1995) allow the selection of
a sample of 15 objects with detections 
both in mass and in radio emission and 6 objects with upper limits 
in the latter. 

 %%%%%%%%%%%%%%%%%%%%%%%%%%%%%%%%%%% FIG 2new %%%%%%%%%%%%%%%%% 
\begin{figure}
 
\vspace{8cm}
\includegraphics{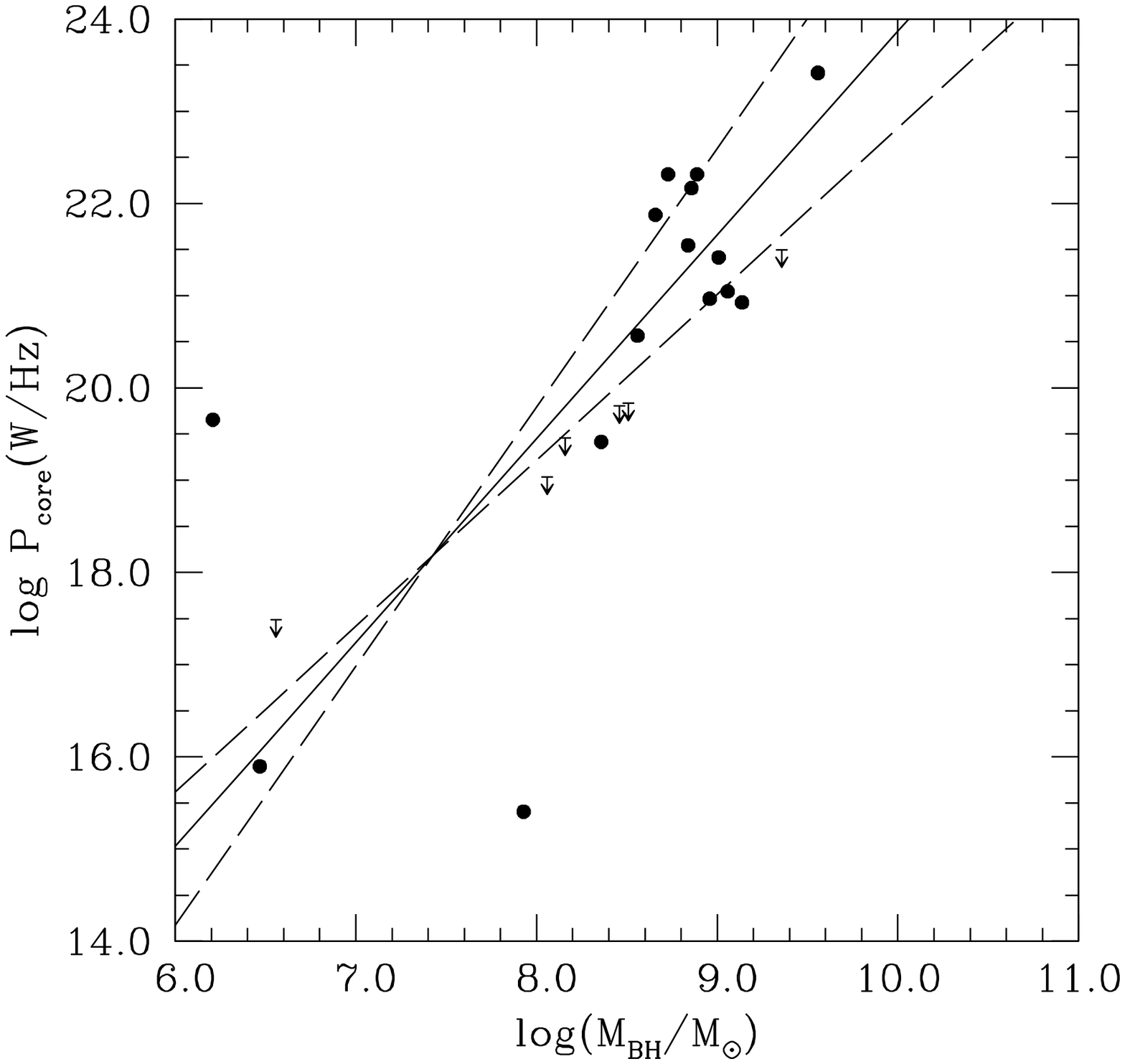}
 \includegraphics{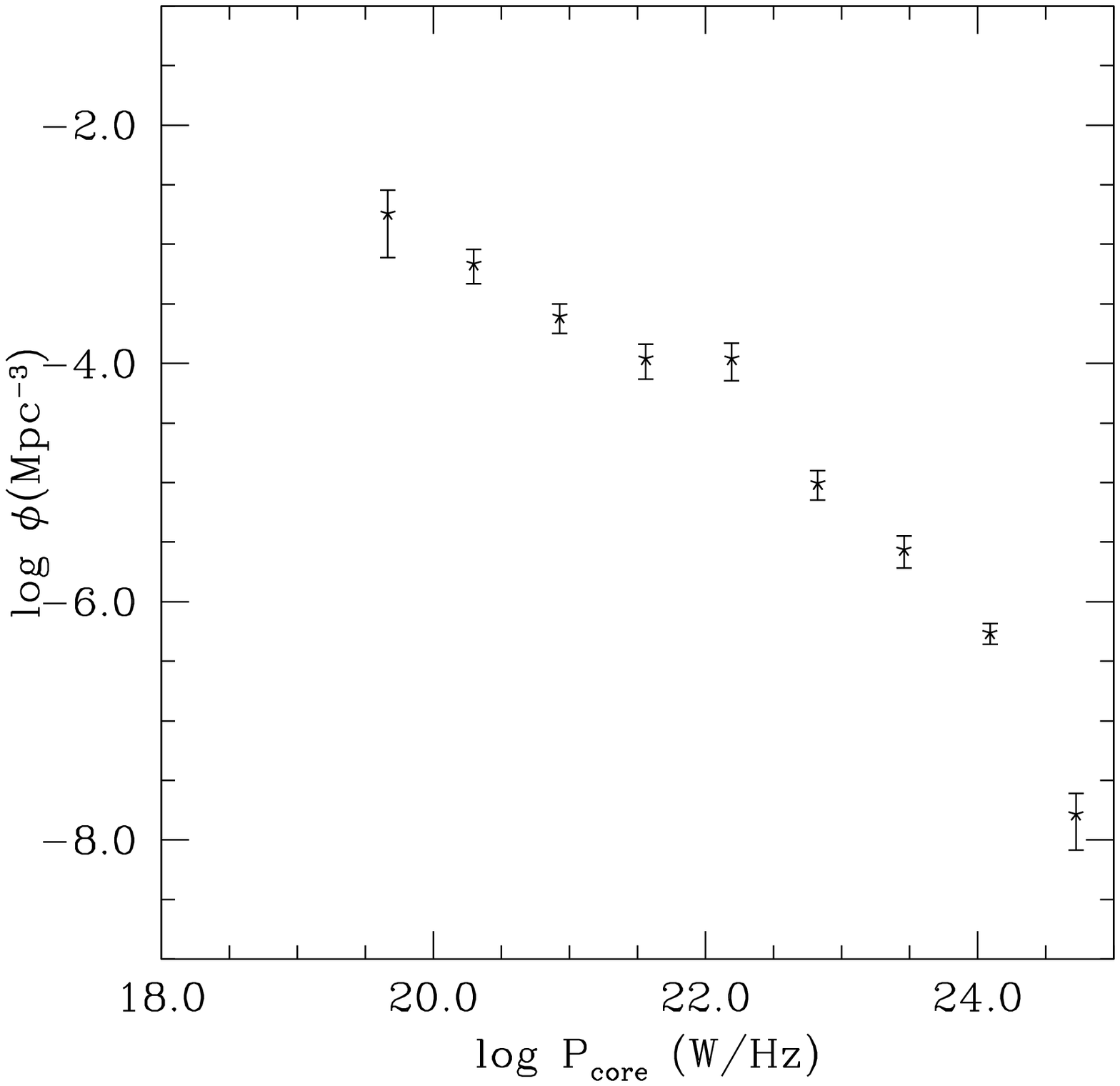}        
         
\caption{{\it Left panel: a)} Mean line regression of  $log \ P_{core}$ vs. $log \ M_{BH}$. 
 (solid line). {\it 
Right panel: b)}
Radio luminosity function (RLF) of the cores of E/S0's Error bars represent 
the statistical errors in the total RLF as estimated by
Sadler et al.\ (1988).
}
\label{fig:fig3}
\end{figure}
%%%%%%%%%%%%%%%%%%%%%%%%%%%%%%%%%%%%%%%%%%%%%%%%%%%%%%%%%%%%%%%%%%

The correlation and the linear regressions have been computed with the 
package ASURV (LaValley, Isobe \& Feigelson 1992). The 
probability of correlation by chance turns out to be $P\lesssim 0.4 \%$.
The best estimate of the regression is:
\begin{equation}
log \  P_{core}^{5GHz} \simeq 19.5 + 2.2_{-0.35}^{+0.55}\ log \ \Big( {M_{BH} \over 
{10^8 M_\odot}}\Big),
\end{equation}
with a slope close to the theoretical inferences for the 
exponent $\alpha$ (figure) 2a. It is worth noticing that the same value of the slope
has been claimed also by McLure et al.\ (1998).

In order to obtain the MF of BHs in nearby galaxies, we need 
to convolve equation (2) with the radio luminosity function
of the nuclei of quiescent spheroidal galaxies. 
Sadler et al.\ (1988) surveyed  114 nearby E and S0 radio--quiet
galaxies at 5 GHz to 
search for weak radio sources and computed
the total power function $\phi(P_{t}) dlog \  P_{t} $ at 5 
GHz down to $log \  P_t({\rm W/Hz})=19.6\ $.
More recently Slee et al. (1994) observed a large fraction of objects
from the same
sample at an  higher angular resolution; this  has  provided the $P_t \ vs. \
P_{core}$ relationship:
 $ log \ P_{core}= 19.5+ 0.78 \ ( log\ P_t -19.5)$
for $log\  P_t\geq 19.5$ and $log \ P_{core}=log \ P_t $ at
lower powers. We can then convert the radio total power
function of Sadler et al.\ (1988) into the core  power funtion (PF) (figure 2b) 
suitable for the present study. 

By using equation (2), we pass from the radio core PF to 
the MDO/BH mass function (hereafter RMF). The result
is shown in figure (3a) alongside with its uncertainties. 
Let us notice that for obtaining the BH mass function,  the 
available statistics can only roughly estimate
the uncertainties related to systematic errors in the procedure.
While the errors
in density represent only the statistical errors of the radio LF,
we estimate the errors in mass from the relationship 
$log \ M_{BH}$ versus  $log  \  P_t$, taking into account 
its uncertainties in the normalization and in the slope. 
The errors turn out to be about $\pm$0.2 dex.
It is worth noticing that the core radio luminosity
function exhibits a gentle slope and no exponential decline at least 
for powers $P\lesssim 10^{24}$ W/Hz. This ensures that
the  possible scatter in the $log \  P_{core} $ versus $log \ M_{BH}$
relationship does  not significantly affect our
results. 

Franceschini et al.\ (1998) estimated the MF using the
total radio luminosity  function of the E and S0 galaxies
and the correlation of the total power with
the BH mass $P_t\propto M_{BH}^{2.66}$. The resulting MF
stays significantly below our estimate
for $M_{BH}>2\times 10^8 \ M_{\odot}$ and the predicted total mass density
is smaller by a factor of about 3, mainly because of the steeper adopted
relationship $P_t-M_{BH}$.

In figure 3a we also plot the MF 
determined  via the optical LF. The comparison should be done
with the lower curve,
which refers to the E/S0 galaxies as the RMF.
Although the two MFs are not completely independent, 
nonetheless, their agreement  supports the reliability
of both estimates. The dependence of the two methods on
the distance scale is quite
similar and thus equivalent matches can be found also for $h=0.5$.
It is worth noticing that  the OMFs derived by  using the distribution
of $log \ x$ proposed by M98 (see eq. 1-2 and figure 1b) 
cannot  be reconciled with the RMF.

\section{The local mass function of the dormant Black Holes}
The mass density associated to luminosities higher
than $L$ is :

\begin{equation}
\rho(>L)=\
{k_{bol}\over \epsilon c^2 H_o} \int {1+z\over (1+z)^3
(1+\Omega z)^{1/2}} dz \int_L ^{L_{max}} L n(L,z)dL
\end{equation}

where $L$ is the luminosity in a certain band and $k_{bol}$ is the
corresponding bolometric correction, $\Omega$ is the density parameter
and $n(L,z)$ is the luminosity function at redshift $z$. As for 
the conversion efficiency of rest mass of the accreted matter into
energy, we adopt
$\epsilon=0.1$, unless otherwise stated. Putting
$L=L_{min}\sim 10^{44}\ erg/s$ (e.g Pei, 1995), we get the local total 
mass density $\rho_{BH}$.
As shown by Soltan (1982)
this quantity can be written in term of source counts and is independent 
of the cosmological parameters.

The mass function of the matter accreted onto AGN
during their activity can now be derived by means of the 
large amount of data  available on  spectra and on
the evolution of the luminosity function. 
The increasing evidence of the presence of MDOs (i.e. inactive
BHs)  in a large number of galaxies 
with significant spheroids 
favors the scenario whereby  the nuclear activity 
is a relatively short phase occurring in a large fraction of galaxies.
The continuity paradigm is then excluded. 
 
 %%%%%%%%%%%%%%%%%%%%%%%%%%%%%%%%%%% FIG 3new %%%%%%%%%%%%%%%%% 
\begin{figure}
 
\vspace{8cm}
\includegraphics{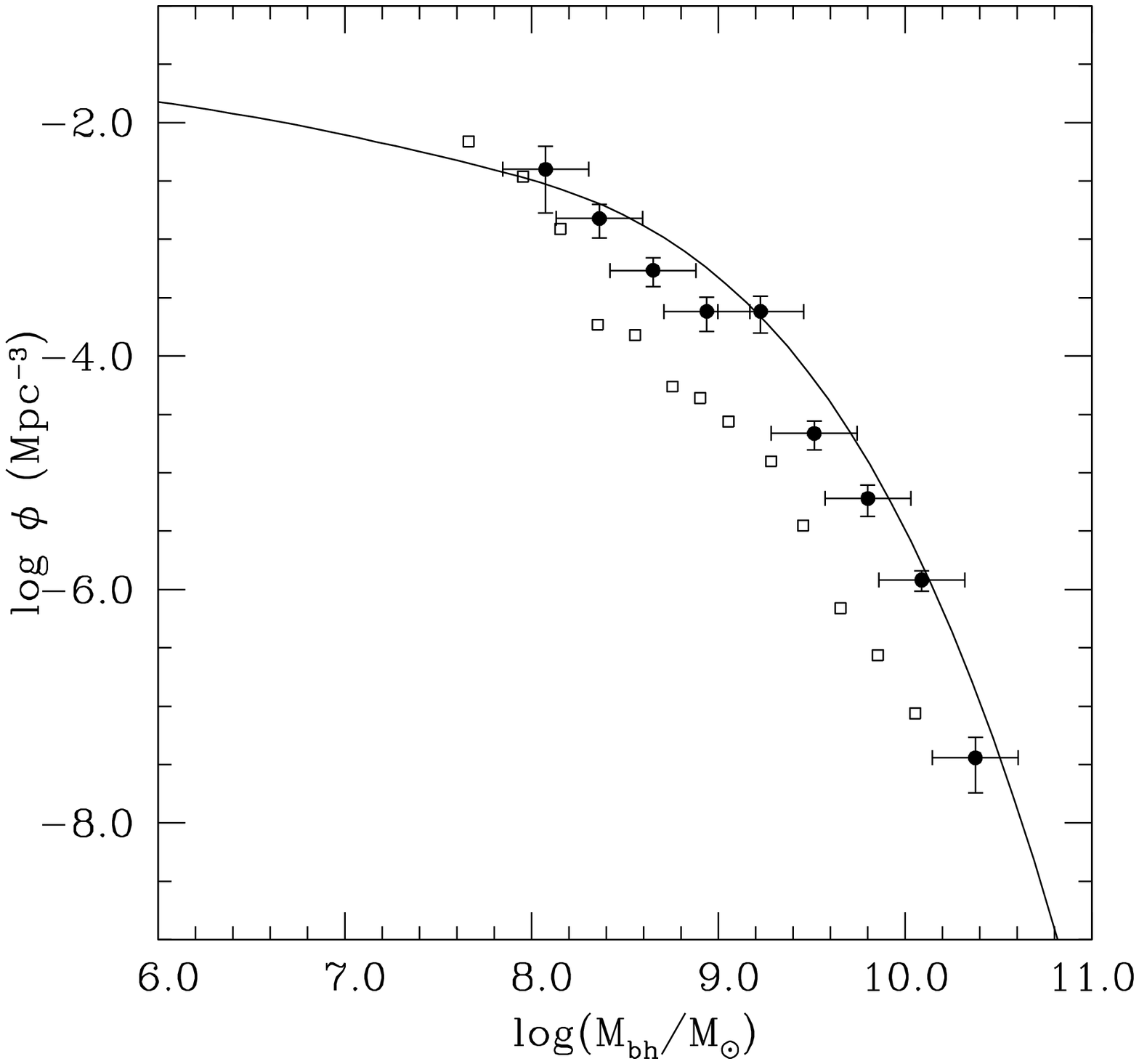}
\includegraphics{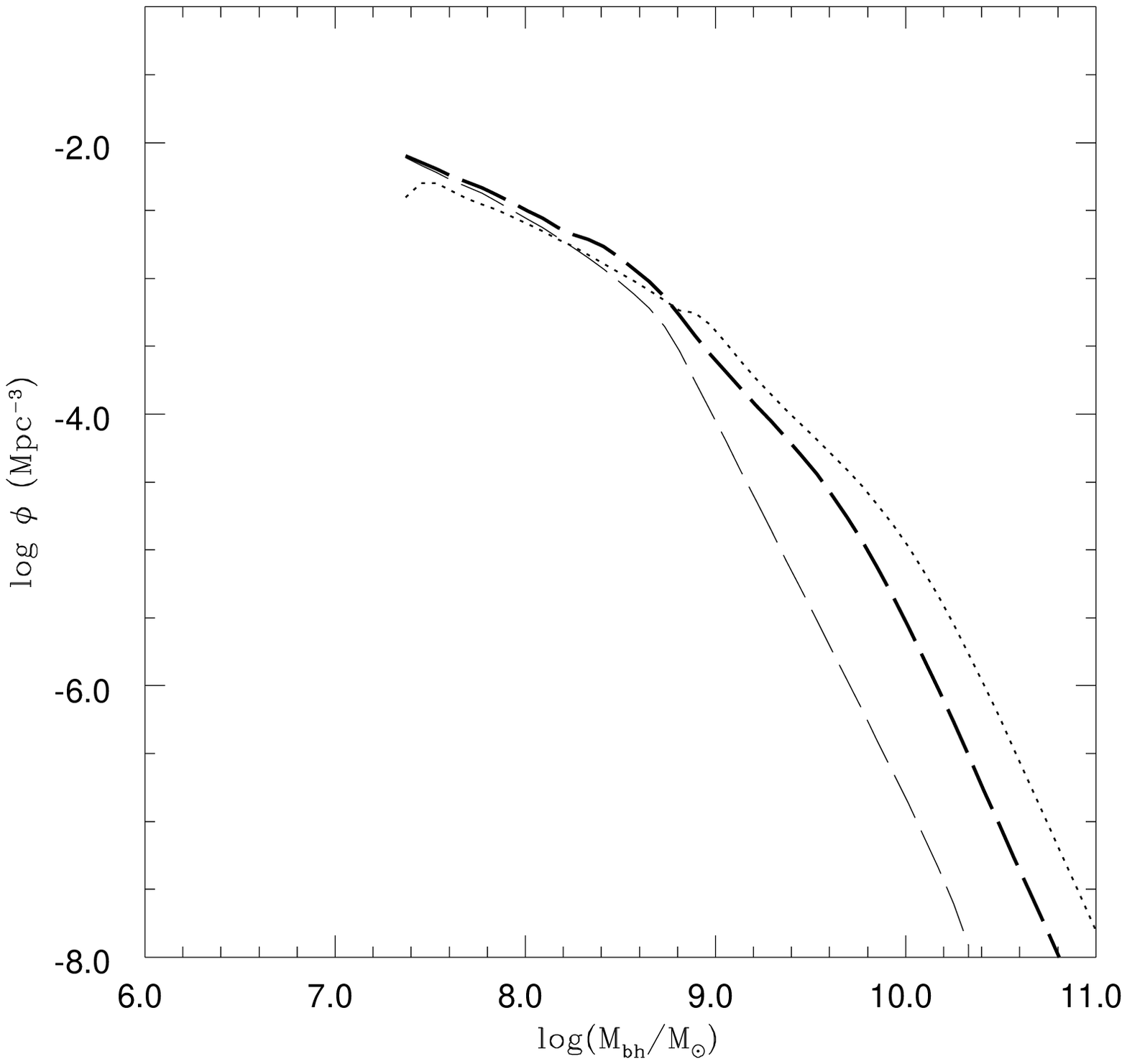}        
\caption{ {\it Left Panel: a) } The  elliptical's OMF mass function
  compared to the mass function, $\phi_{RMF}\ dlog \ M_{BH}$,  derived from the 
E/S0 cores   radio PF (filled circles). Also shown the RLF derived by 
  Franceschini et al. (1998) {\it Right Panel: b)} The total mass function of
   the relic BHs, $\phi_{AMF}\ dlog \ M_{BH}$,  derived from the 
past activity of AGN's  
(thick dashed line);  also shown the contribution from the 
AGN's originating the HXRB (thin-dashed line) and   the total mass function 
for $h=1/2$ (dotted  line). }

\label{fig:fig5 }
\end{figure}
%%%%%%%%%%%%%%%%%%%%%%%%%%%%%%%%%%%%%%%%%%%%%%%%%%%%%%%%%%%%%%%%%%

In the hypothesis of  single short events
it is likely that the AGN are observed in the highest state of activity,
then,  we can evaluate the mass of the BH associated to a luminosity $L$,
 (hereafter the bolometric luminosity)
through the assumption $L=\lambda L_{Edd}$  and $L_{Edd}=M_{BH}c^2/t_E$. 
The constraint for the latter can be obtained as it follows:
if we assume  $\lambda$ constant  during the accretion, 
we get:
\begin{equation}
M_{BH}(t_{obs})=M_{BH}(t_{in})\exp \Big({\lambda \tau_o \over \epsilon
t_E}\Big),
\end{equation}
and
\begin{equation}
L(t_{obs})={\lambda c^2 \over t_E }M_{BH}(t_{obs})=
{\lambda c^2 \over t_E }M_{BH}(t_{in})
\exp\Big({\lambda \tau_o \over \epsilon
t_E}\Big),
\end{equation}

where $t_E= (\sigma_T c )/(
4\pi m_p) \simeq 4\times 10^8$\ yr, and $t_{in}$ and
$t_{obs}$ are the initial time of the bright phase
and the observing time respectively,
and $\tau_o=t_{obs}-t_{in}$. Then, by requiring  
$M_{BH}(t_{obs})\gg M_{BH}(t_{in})$ we get the constraint: $\tau_o\geq
(2 \epsilon t_E) / \lambda$.

Following equation (4), we  write the QSO/AGN luminosity function
 $\psi(L^\prime)$ as

\begin{equation}
\int _{0}^{ L^\prime} {d\psi (>L)\over dL} dL=
\int_{-\infty}^{log \ M_{a}^\prime(L^\prime )} M_{a} \phi_{AMF} (M_{a}) dlog\ M_{a}
\end{equation}

where $\phi_{AMF} (M_{a})$ is the mass function of the accreted mass,
 $M_{a}=M_{BH}$. Since
  $\lambda$ is likely a weakly increasing
function  of the luminosity, let us then
set:
\begin{equation}
\lambda(L) =L/L_E=10^{\gamma (log \ L-49)}\ \ \ ; \  \ \gamma
\simeq 0.2
\end{equation}

that follows the   findings that
the most luminous QSOs radiate at about the Eddington limit,
while low luminosity AGN, $L\sim 10^{44} \ erg/s$, show 
 $\lambda=L/L_E\sim 0.1-0.05$
 (see Padovani (1989) and Wandel (1998)). 
Then, from equation (6) and (7) the MF
of the relic BHs can be written as:
\begin{equation}
\phi (M_{BH}) dlog \ M_{BH}={ln(10)\over (1-\gamma)}{\psi (>L)\over M_{BH}} 
{dlog \ (\psi(>L))\over 
dlog \ (L)} \ dlog \ M_{BH}  
\end{equation}

Next step is to insert
into equations  (3) and (8) the information on the 
luminosity function and its evolution. For the optically selected
AGN we adopt
the luminosity function and its cosmic evolution for the B--band 
given by Pei (1995), which is well defined at least
up to $z\sim 3.5 $. 
The bolometric correction for the B--band
has been taken $k_{bol}(B)=13$, on the basis of the spectra reported by
Elvis et al (1994). The total mass density turns out to be
$\rho_{BH}\simeq 2.0 \times 10^{5}\ M_{\odot}/Mpc^{3}$, close to the 
estimate of Chokshi \& Turner (1992), who however adopted
$k_{bol}(B)=16.5$.

The mass density of matter accreted on massive BHs powering
optically selected AGN  has been often used as a reference for the total
mass density in BHs. However, as pointed out by
Granato et al.\ (1997), the total amount of the  matter accreted onto
AGN should include 
that required to produce the 2-50 keV background (HXRB).
In fact,  the surveys in the optical and in the soft 
X--ray bands tend to select expecially type I AGN and QSOs (Hasinger 1998) 
and to lose a significant
fraction of absorbed active objects, the so called type II AGN. 

Setti and Voltjer (1989) suggested that these AGN  yield most of
the HXRB (let us remind that type I AGN are well known to be  minor
contributors to the HXRB).
A direct evidence comes from  Fiore et al (1998), that,  in a 
survey with BeppoSAX
in 5-14 keV band, found about 150 sources corresponding to a surface density
of 20 objects/sq degree at the flux limit 
$F\simeq 6 \times 10^{-14}$ \ erg/s/cm$^{2}$. So far, there have been  
optically identified only 9 sources, 5 of which  turn out to 
be QSO's and 4 ``type II'' AGN. The surface density  implied 
by these detections
 can explain 
about 40$\%$ of the HXRB. 
 
A suitable  estimate of the mass density underlying the
intensity of the HXRB between 2-50 keV can be obtained
through the relationship
\begin{equation}
\rho_{BH}={E_{bol}\over \epsilon c^2}\simeq {k_{bol} 
\over \epsilon c^2}{(1+z_e)\over
k(z_e)} {I\over c}.
\end{equation}
The intensity in the 2-50 keV  band is
$I\simeq 1.9 \times 10^{-6}$\ erg \ cm$^{-2}$\ sec$^{-1}$, 
$k_{bol}\simeq 14.5$ is the bolometric correction for 2-50 keV band 
(see Elvis et al.\ 1994; Bassani et al. \ 1998),  $z_e$ is the
{\it effective} emission redshift, which can be assumed
$z_e\sim 1$ and $k(z_e)$ is the k--correction for $z=z_e$. 
The total associated mass density amounts to $\rho_{BH}
\sim  3-5 \times 10^5 \ M_{\odot}/Mpc^{-3}$, where the uncertainty
reflects the uncertainty on $z_e$, which disappears once
$n(L,z)$ is known.

Direct information on $n(L,z)$ for
objects selected in hard X--ray bands is still scarce.
However, assuming that the HXRB is chiefly
ascribed to type II AGN and that the unified schemes are basically
valid, then the soft X--ray observations and the shape of the HRXB
significantly constrain the luminosity function and the evolution
of the type II AGN. In order to compute the mass function, 
we use the models proposed
by Comastri et al.\ (1995) and by Celotti et al (1995) to
reproduce the HXRB.
Comastri et al (1995) used a LF in the 0.3--3.5
keV band, while Celotti et al (1995) one referring to the 2-10 keV band. We 
adopted $k_{bol}=25$ and  $k_{bol}=38$ respectively,
on the basis of the
spectra presented by Elvis et al.\ (1994). 
Both models assume luminosity evolution $L(z)\propto (1+z)^{k}$ with
$k\sim 2.6-3.0$ and a similar redshift cutoff $z\sim 2-3$ beyond which
the evolution stops. The constraints imposed by the shape of the HXRB
and 
by the LF and the evolution of the soft X--ray selected AGN, compel
the  redshift distribution to be quite similar for both models. Let us
notice that 
the shape of the mass function and the
mass density $\rho_{BH}(X-ray)\simeq 4.2 \times 10^{5}\ M_{\odot}/Mpc^{3}$
results  comfortably very similar in both models.

In figure 3b we show the mass function derived for $\lambda=1$ and 
$\lambda$ given by eq (7).  
The corresponding total mass density is $\rho_{BH}\simeq 6.5 \times 10^{5}
\ M_{\odot}/Mpc^{3}$.
It is apparent
that optically selected AGN dominate at $M_{BH}> 10^9$\ M$_{\odot}$, while
BHs associated to the X--ray absorbed AGN dominate
the MF at $M<10^8$\ M$_{\odot}$. 
Unlike the total mass density, the mass function of the accreted matter
(hereafter AMF) does depend on the cosmological parameters.
However, as shown it figure (3b), if we decrease $h$  down to $0.5$, 
the AMF, 
for $M_{BH}> 5\times 10^9$, changes slightly .

The local LF of AGN in the X--ray bands do not extend 
below $\ sim 4\times 10^{42}$ erg/s so that  
our estimate includes only objects 
with $L >10^{44}$ erg/s and with $M_{BH}>10^7$  M$_{\odot}$,
assuming $\lambda \simeq 0.1$ at low luminosities. In order to probe
masses below this limit we should include the, presently not very well-known
 contribution from 
low luminosity active nuclei, such as the LINERS.

In figure (4a) we compare the MF derived from the past activity,
assuming $\lambda=1$ and $\lambda=0.2$, to the mass functions derived 
from the optical and radio LFs.
The slopes are quite different and the agreement may be considered
acceptable only in the range $10^8 \  M_\odot - 10^{10} \ M_\odot$.
This
suggests that the assumption of a constant $\lambda=L/L_E$ is not adequate. 

Instead, by using eq. (7), which is motivated on observational grounds,
the agreement is excellent.
In detail, since the total mass density is constant, the individual 
masses of low luminosity objects increase while their number decreases,
and the AMF of the inactive BHs turns out in good agreement with the MF 
of the local MDOs (see figure (4b)).
This dependence may be suggestive of recurrent activity confined to
 low mass objects.

 %%%%%%%%%%%%%%%%%%%%%%%%%%%%%%%%%%% FIG 4new  %%%%%%%%%%%%%%%%% 
\begin{figure}
\vspace{8cm}
\includegraphics{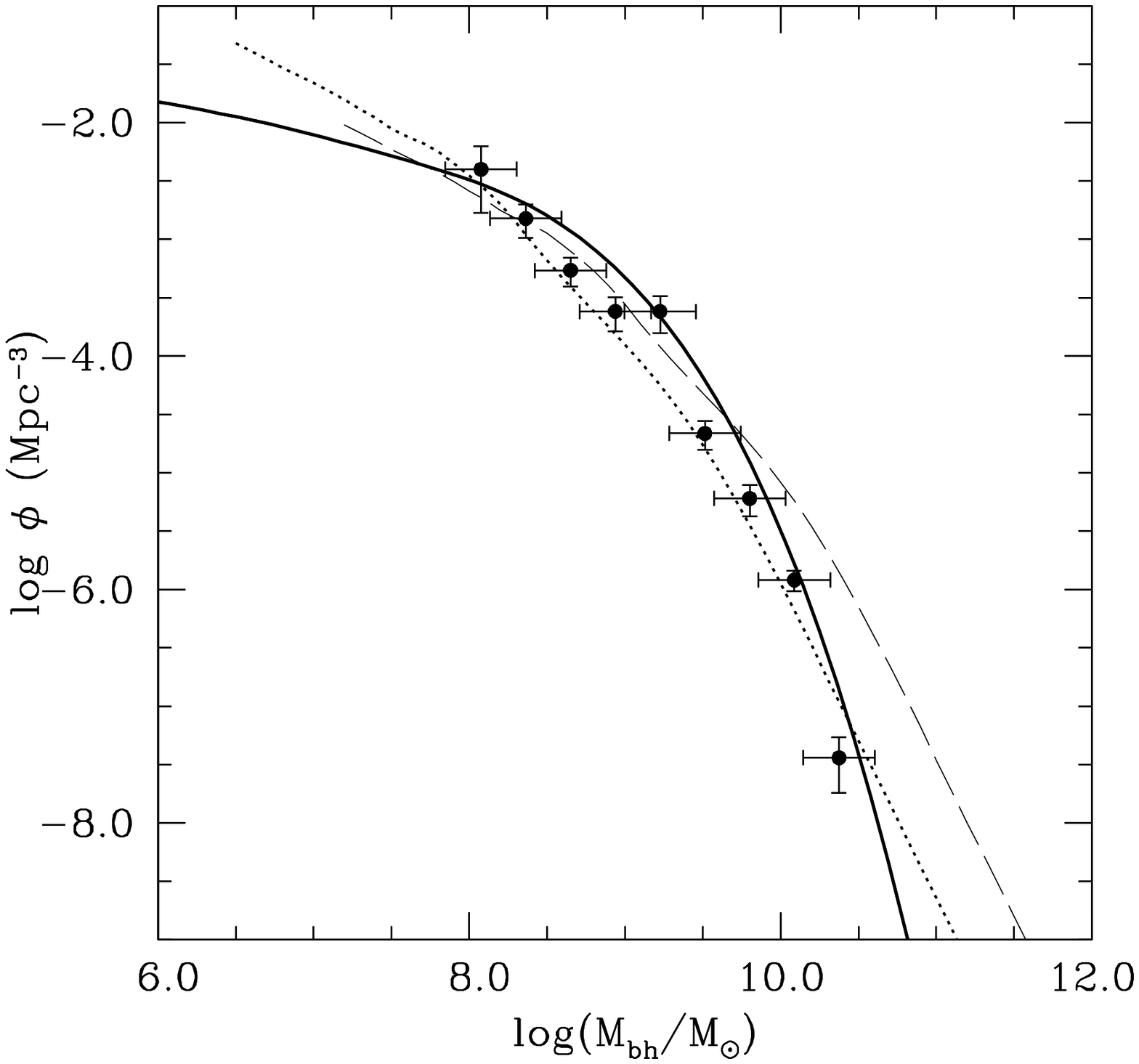}
\includegraphics{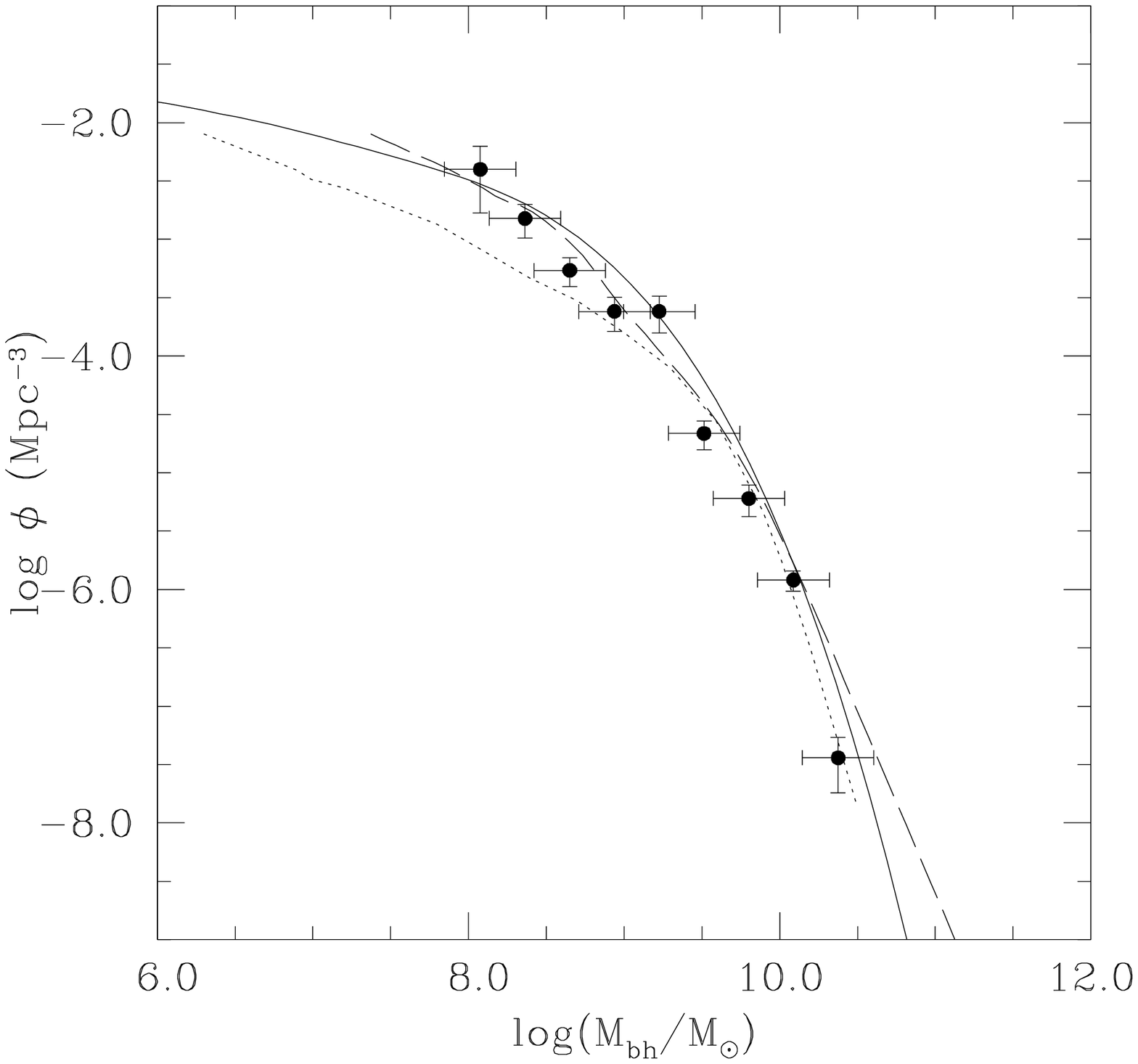}

\caption{{\it Left panel: a)}
The AMF  derived by
assuming  $\lambda=1$  (dotted line) and $\lambda=0.2$ (dashed line).
In both panels the solid line and filled circles are  defined as in 
fig. (3a).
 {\it Right panel:
 4b) } The AMF  derived according to eq. (7) compared with the
 MF of the relic BHs by Cavaliere and Vittorini (1998) (dotted line).
 }
\label{fig:fig7 }
\end{figure}
%%%%%%%%%%%%%%%%%%%%%%%%%%%%%%%%%%%%%%%%%%%%%%%%%%%%%%%%%%%%%%%%%%

\section{Discussion and conclusions}

The agreement
found between the mass functions derived from investigations
on MDOs resident in local galaxies and the mass function
of the BHs inferred from the past activity of AGN is based on
 very simple and sound hypotheses: {\it i)}
the nuclear activity is a single short event; {\it ii)} the spectra 
of the AGN do not greatly depend on  redshift; 
{\it iii)} the mass-radiation conversion efficiency
of accretion $\epsilon\simeq 0.1$; {\it iv)} the HXRB is produced by
absorbed AGN;
{\it v)} 
$\lambda=L/L_E$ is a weak increasing function of the luminosity.

The inclusion in the estimate of the mass deposited in BHs by
the activity related to the HXRB is mandatory, 
independently of the specific model of HXRB 
one adopts. In fact, the optically selected
objects contribute a minor fraction $\lesssim 20\%$ of the HXRB,
whose  observed intensity implies $\rho_{BH}(X-ray)
\sim  3-5 \times 10^5 \ M_{\odot}/Mpc^{-3}$. 

Let us comment that the idea that  ADAF accretion would be 
 responsible for 
the HXRB (Di Matteo and Fabian, 1997,
 see also Haenhelt et al 1998), which  implies a  quite larger 
 $\rho_{BH}(X-ray)$  and a different AMF with respect to the
 one we have derived,
runs against the evidence from the  new hard X--ray surveys that
 have found that 
 obscured objects  are responsible for the 
emission of at least 20 $\%$ of the HXBR, yet at a rather bright
flux limit (Fiore 1998).

The MF computed using  the 
distribution of the ratio $M_{BH}/M_{sph}$ obtained by M98  
is in contrast with that
derived from  the correlation found between MDO mass and radio 
core power. If we neglect this fact,
the OMF predicted by means of  the log normal distribution of the 
$M_{BH}/M_{sph}$ ratio of M98 can match the AMF  by assuming an efficiency 
$\epsilon\sim 0.02$, a factor 5 less than the  standard
value. Alternatively, the bolometric corrections should be increased by the 
same unlikely factor.

   Large mass densities could  be explained by  
assuming that the mass accreted 
during bright phases that shows up directly in the optical counts or 
in the integrated HXRB is 
only a small fraction of the total mass of the BHs; in this case
most of the mass must be accreted in a silent phase (in  the optical
and in the X--ray bands).However,  If the silent phase occurred before the bright 
one, the problem of BH formation at early time would become
more intricate and $\lambda<< 1$ is required at all luminosities. 
If it is postponed, then extreme obscuration is required.

The dependence 
on $H_o$ of the mass functions is interesting. The shape of the AMF  
depends slightly  on $H_o$ while the corresponding  total
mass density of the accreted matter is independent of $H_o$.
On the other hand,
the mass functions derived from the local
radio power function and optical LFs strongly depend on $H_\circ$,
with $M_{MDO}\propto h^{-1}$ and $\Phi \propto h^{3}$ while 
the mass density in dormant BHs $\rho_{BH}
\propto h^2$. With the present uncertainties in the 
derivation of the mass functions it is not possible
to discriminate among different values of the Hubble constant.
In order to counterbalance
the decrease by a factor of about 2 of the total luminosity 
density due to the change from
$H_{\circ}=70 km/s \ Mpc^{-1}$ to $H_{\circ}=50 km/s \ Mpc^{-1}$, the average ratio
$M_{BH}/M_{sph}$ must be increased by the same factor.
Thus the match between
OMF and AMF is obtained   with $log \ x_{\circ}=-2.35$
and $\sigma =0.3$ in eq (1). The agreement with the
RMF is obtained assuming $
log  \ P_{core}^{5GHz} \simeq 19.2 + 2.0\ log \ (M_{BH}/ 10^8\ M_\odot) $,
which is  a good fit to the  data.

However it is conceivable that in the near future additional
high resolution photometric, spectroscopic and radio observations
and very deep hard X--ray surveys
will allow to significantly reduce the uncertainties
in the estimate of OMF, RMF and AMF.
At that point the possibility 
of balancing changes of the distance scale acting 
on some parameters  will be interestingly restricted.

The activity patterns have been investigated by several authors
(e.g. Cavaliere and Padovani 1988;
Small and Blandford 1992; Haehnelt and Rees 1993;
Cavaliere and Vittorini 1998). 
 Cavaliere and Vittorini (1998) propose a scenario in which  the energy output,
 at
early times are governed by hierarchically growing 
environment, while later,  the fall in the average 
luminosity is related to intermittent accretion governed by 
galaxy--galaxy interactions. Their prediction
of the  local MF,  reported in figure 4b, 
is  below our 
estimate  in the mass 
range $10^7 \ M_\odot \lesssim M_{BH}\lesssim 10^9 \ M_{\odot}$.
 The derived 
total mass density is $\rho_{BH}\simeq 3.2 \times 10^5\ M_{\odot}
\ Mpc^{-3}$, falls short by a factor two  
the total mass accreted onto type I and type II AGN.
The difference may be due to the fact that BHs in early type spirals
are not considered by Cavaliere \& Vittorini (1998).

In conclusion  our analysis strongly supports  that the nuclear activity 
is in general a single event, with  bolometric luminosities close 
to the Eddington luminosities $\lambda=L/L_{Edd}$ but increasing increase from
$\lambda \sim 0.1$ for  $L
\sim 10^{44}\ erg \ sec^{-1}$ to 
$\lambda \sim 1$ for $L
\gtrsim 10^{48} \ erg \ sec^{-1}$, as found also
by independent studies (see e.g. Padovani
1989). This suggests that recurrent activity occurs in low
luminosity objects and/or that their 
accretion flows are supply--limited and cannot sustain Eddington 
luminosities.

In our framework  type II AGN of small and moderate mass
are responsible for most of  the HRXB. These objects can 
reside in early type galaxies but also
in spirals with still significant spheroidal
component (Sa/Sab). 
On the other hand, bright objects such as QSOs at
significant redshift should be hosted preferentially by E/S0
galaxies (or by their precursors). This has been recently confirmed
by a  study of the host galaxies of QSOs with $M_R> -24$ (McLure et al.\ 
1998). These author found that a large fraction of the host galaxies
are massive elliptical galaxies.
This facts add complexity 
to the problem of the AGN evolution. Not only coordination
is required to mimic the luminosity evolution,
but also a kind of coordination 
is required to pass from the high luminosity, high redshift
optically selected AGN, preferentially hosted in E/S0 galaxies, to
the low luminosity objects , such as the local Seyfert galaxies,
hosted preferentially in Sa/Sab galaxies.
The discussion of these aspects in relation with the problem
of BH formation will be presented in a subsequent paper
(Monaco et al 1998).
 
\medskip

\section{ACKNOWLEDGMENTS}
We thank A.\phantom{ }~Cavaliere,
F. La Franca and  G. Fasano for helpful discussions  and J. Magorrian for 
insightful and clarifying comments. We thank also the anonymous referee for comment
that have helped to better present our  results. P.M. has been supported  
 by the EC Marie Curie TMR contract ERB FMB
ICT961709.

\bigskip
\bigskip
\bigskip

\centerline{REFERENCES}
\vglue 0.2truecm

\ref {Abramowicz M., Czerny B., Lasota J.P., Szuszkiewicz E.,
1988, ApJ 332, 646}

\ref {Antonucci R., 1993, ARA\&A, 31, 473

%\ref {Binney J., Mamon G.A., 1982, MNRAS, 200, 361}

%\ref {Boyle B.J., Shanks T., Peterson B.A., 1988, MNRAS, 232, 431}

\ref {Bassani, L., Cappi, M., Malaguti, G.: 1999,  astro-ph 9901144}

\ref {Cavaliere A., Padovani P., 1988, ApJ, L33}

\ref {Cavaliere A., Vittorini V., 1998, in The Young Universe;
Galaxy Formation and Evolution at Intermediate and High Redshift,
eds. S. D'Odorico, A. Fontana, E. Giallongo, 
Astron. Soc. Pac. Conf. Ser. 146, 26}

\ref {Chokshi A., Turner E.L., 1992, MNRAS, 259, 421}

\ref {Celotti A., Fabian A.C., Ghisellini G., Madau P., 1995,
MNRAS, 277, 1169}

\ref {Comastri A., Setti G., Zamorani G., Hasinger G., 1995, A\&A,
296, 1}

\ref {Di Matteo T., Fabian A.C., 1997, MNRAS, 286, 393}

\ref {Di Matteo T., Fabian A.C., Rees M.J., Carilli C.L., Ivinson
R.J., 1998, astro--ph/9807245}

\ref {Efstathiou G., Ellis R.S.,Peterson B.A., 1988, MNRAS,
232, 431}

\ref {Elvis M., Wilkes B.J., McDowell J.C., Green R.F., 
Bechtold J., Willner S.P., Oey M.S., Polomski E.,
Cutri R., 1994, ApJS, 95, 68}

\ref {Fiore et al 1998, Nature, submitted }

\ref {Ford H. C., Tsvetanov Z.I., Ferrarese L., Jaffe W.,
1997, preprint astro-ph 9711299, in 
Proceedings IAU Symposium 186, Kyoto, August 1997, D.B. Sanders,
J. Barnes, eds., Kluwer Academic Pub.}

\ref {Franceschini A., Vercellone S., Fabian A.C., 1998, preprint
astro-ph 9801129}

\ref {Fukugita M., Hogan C.J., Peebles P.J.E., 1997, preprint
astro-ph 9712020}

\ref {Geller M. J., Kurtz M.J., Wegner G., Thorstensen J.R.,
Fabricant D.G., Marzke R.O., Huchra J.P., Schild R.E.,
Falco E.E., 1997, , 114, 2205}

\ref {Granato G.L., Danese L., Franceschini A., 1997, ApJ, 486}

\ref {Haehnelt M.G., Rees M.J., 1993, MNRAS, 263, 168}

\ref{ Hasinger G., 1998, Astr. Nachtr, 319, preprint astro-ph 9712342}

\ref {Heyl J., Colless M., Ellis R.S., Broadhurst T., 1997,
MNRAS, 285, 613}

\ref {Ho L.C., 1998, preprint astro-ph 9803307, in Observational
Evidence for Black Holes in the Universe, ed. S.K. Chakrabarti,
Kluwer Academic Pub.}

\ref {Kent S.M., 1985, ApJ Suppl., 59, 115} 

\ref {Kodaira, K., Watanabe, M., Okamura S., 1986, ApJ Suppl.,
62, 703}

\ref {Kormendy J., 1993, in The Nearest Active Galaxies, eds. J. Beckman,
L. Colina, \& H. Netzer (CSIC Press, Madrid), 197}

\ref {Kormendy J., Richstone D., 1995, ARA\&A, 33, 581}

\ref {LaValley M., Isobbe, T. , Feigelson E.D., 1992, BAAS,
24, 839}

\ref {Loveday J., Peterson B.A., Efstathiou G., Maddox S.J.,
1992, ApJ, 390, 338}

\ref {Magorrian J., Tremaine S., Richstone D., Bender R.,
Bower G., Dressler A., Faber S.M., Gebhardt K., Green R.,
Grillmair C., Kormendy J., Lauer T.R., 1998, , 115, 2285}

\ref {Mahadevan R., 1997, ApJ, 605}

\ref {McLure R.J., Dunlop J.S., Kukula M.J., Baum S.A.,
O'Dea C.P., Hughes D.H., 1998, preprint astro-ph/9809030}

\ref {Narayan R., Yi I., 1995a, ApJ, 444, 231}

\ref {Narayan R., Yi I., 1995b, ApJ, 452, 710}

\ref{Padovani P., 1989, A \& A, 209, 27}

\ref {Pei Y.C., 1995, ApJ, 438, 623}

\ref {Rees M.J., Begelman M.C., Blandford R.D., Phinney
E.S., 1982, Nat., 295, 17}

\ref {Rees M.J., 1996, in Black Holes and Relativity, ed. R. Wald,
Chandrasekhar Memorial Conference, Dec. Astroph 9701161}

\ref {Rees M.J., Fabian A.C., 1995, MNRAS, L55}

\ref {Sadler E.M., Jenkins C.R., Kotanyi C.G., 1989, MNRAS, 240, 591}

\ref {Sadler E.M., Slee O.B., Reynolds J.E., Roy A.L., 1995,
MNRAS, 276, 1373}

\ref{Salucci P. et al.\, 1998, submitted MNRAS}

\ref {Schechter P.L., Dressler A., ,1987, 94, 563}

\ref {Setti G., Woltjer L., 1989, A \& A 224, L21

\ref {Simien F., de Vaucouleurs G., 1986, ApJ, 302, 564}

\ref {Slee O.B., Sadler E.M., Reynolds J.E., Ekers R.D., 1994,
MNRAS, 269, 928}

\ref {Small T. A. \& Blanford A. D., 1992, MNRAS, 259, 725}

\ref {Soltan A., 1982, MNRAS, 200, 115}

\ref {van der Marel R.P., 1991, MNRAS, 253, 710}

\ref {van der Marel R.P., 1997, preprint astro-ph 9712076, in 
Proceedings IAU Symposium 186, Kyoto, August 1997, D.B. Sanders,
J. Barnes, eds., Kluwer Academic Pub.}

\ref {van der Marel R.P., 1998, preprint astro-ph 9806365, submitted
to Ap.J.}

\ref {Wandel A., 1998, astro-ph/9808171 }

\ref {Wrobel J.M., 1991, AJ , 101, 127}

\ref {Wrobel J.M., Heeschen D.S., 1991,AJ , 101, 148}

\ref{Young P. 1980, ApJ, 242,1232}

\ref {Zucca E., et al.\ , 1997, A\&A, 326, 477}

\vfill\eject

\end{document}